\documentclass{article}

\usepackage{tabularx,booktabs}
\usepackage{lineno}
\usepackage[margin=0.9in]{geometry} 
\usepackage{amsmath,amsthm,amssymb}
\usepackage{multicol}
\usepackage{graphicx}
\usepackage{caption}
\usepackage{booktabs}
\usepackage{multirow}
\usepackage{siunitx}
\usepackage{xcolor}
\usepackage{authblk}
\usepackage{natbib}

\usepackage{url}

\title{A Two-Stage Deep Learning Detection Classifier for the ATLAS Asteroid Survey}
\author[1]{Amandin Chyba Rabeendran}
\author[2]{Larry Denneau}
\affil[1]{\small Applied Mathematics, Colorado School of Mines, 1500 Illinois St, Golden, CO 80401, amandinchyba@gmail.com}
\affil[2]{\small Institute for Astronomy, University of Hawai`i, 2680 Woodlawn Drive, Honolulu, HI 96822, denneau@hawaii.edu}
\date{\vspace{-5ex}}

\begin{document}

\maketitle

\begin{center}
\end{center}

\begin{abstract}
    In this paper we present a two-step neural network model to separate detections of solar system objects from optical and electronic artifacts in data obtained with the “Asteroid Terrestrial-impact Last Alert System”(ATLAS), a near-Earth asteroid sky survey system \citep{Tonry_2018_2}. A convolutional neural network \citep{Lieu_2019} is used to classify small ``postage-stamp'' images of candidate detections of astronomical sources into eight classes, followed by a multi-layered perceptron that provides a probability that a temporal sequence of four candidate detections represents a real astronomical source. The goal of this work is to reduce the time delay between Near-Earth Object (NEO) detections and submission to the Minor Planet Center. Due to the rare and hazardous nature of NEOs \citep{2015Icar..257..302H}, a low false negative rate is a priority for the model. We show that the model reaches 99.6\% accuracy on real asteroids in ATLAS data with a 0.4\% false negative rate. Deployment of this model on ATLAS has reduced the amount of NEO candidates that astronomers must screen by 90\%, thereby bringing ATLAS one step closer to full autonomy.
\end{abstract}

\begin{multicols}{2}

\section{Introduction}

From the Tunguska impactor in 1908 \citep{foschini2018atmospheric} to the realization by Alvarez et al. that the $\sim$30 km Chicxulub impactor was the likely cause of the K-T extinction event \citep{Alvarez80extraterrestrialcause}, the possibility of an asteroid impact with Earth has been deemed a dangerous threat. In 1998, recognizing such threats, the United States Congress mandated that NASA develop a plan to catalog 90\% of asteroids 1 km and larger whose orbits bring them within 1 AU of the Earth's orbit.

The ``Asteroid Terrestrial-impact Last Alert System" (ATLAS) is a sky-survey system that was made operational in 2016 with the specific goal of detecting Near-Earth Objects (NEOs) with a performant, cost-effective system \citep{Tonry_2018_2}. By balancing unit cost,  processing power, and autonomous operation, ATLAS is able to consistently detect more NEOs within a 0.01 AU distance from Earth than any other asteroid surveys \citep{Heinze_AU}. Candidate new NEOs detected by the ATLAS system are reviewed and confirmed through human screening before being sent to the IAU Minor Planet Center (MPC) for additional followup observations by other facilities. Within the past four years, ATLAS has discovered more than 500 NEOs and submitted over 50,000 NEO observations to the MPC.

ATLAS identifies asteroids by looking for moving sources over four or more survey exposures acquired at the same approximate location on the sky. The coordinates and metadata for a moving object over these exposures form a ``tracklet'', the fundamental unit of information in ATLAS asteroid processing. The ATLAS system detects tens of thousands of known asteroids per night but also generates many hundreds of false tracklets caused by various image contaminants such as variable stars, or optical and electronic artifacts. Candidate unknown tracklets (of NEOs usually) must be screened such that contaminants are not propagated through the global NEO discovery system. Improved automatic identification of bogus tracklets can lead to greater automation within ATLAS and reduce delays in obtaining followup observations. 

Our objective is to train a neural network to separate false tracklets detected as NEO candidates from real NEOs. A primary constraint to the automatic process is retention of all real NEOs. If the number of false tracklets can be substantially reduced, we hope to eventually automatically submit NEO candidates to the MPC, reducing the latency between observation and reporting to the MPC, thereby eliminating most of the need for human screening. 

Deep learning has provided astronomers with new tools to autonomously analyze astronomical sources through predictions based on a trained neural network \citep{baron2019machine}. Convolutional Neural Networks (CNN) represent state of the art accuracy when it comes to image classification \citep{Khan_2020}. We decided to employ a CNN over other deep learning architectures based on its excellent feature extraction capability. Additionally, most other deep learning models that work with 2D imagery such as GANs, R-CNNs, and YOLO use CNNs as a classifier (e.g. the discriminator in GANs) \citep{Schawinski_2017,NIPS2015_5638,redmon2016look}. Therefore a CNN provides a flexible backbone that we can use as a baseline or starting point for future image classification work.

We tested the model on a month of ATLAS tracklets from June 5th to July 5th 2020. We have deployed the model on ATLAS and it is currently being used by astronomers to screen NEO candidates on a daily basis. The deployed model is showing positive results while being improved upon based on astronomer feedback.

\section{ATLAS}

ATLAS telescopes have been in operation since 2017 on two mountaintops in Hawai`i (Haleakal\=a and Maunaloa). Two additional ATLAS telescopes are under construction in the southern hemisphere, in South Africa and Chile. The current two-telescope ATLAS system gathers new data every night on the order of $10^7$ images resulting in over 0.5~TB of raw uncompressed data \citep{Tonry_2018_2}. Approximately $10^4$ full sized images are classified by the ATLAS processing system every night.

Every night when permitted by weather, each ATLAS telescope surveys a set of $\sim200$ pre-defined locations that raster the night sky (each called a ``footprint'') and takes four 30-second exposures at each footprint over a span of $\sim30$ minutes. The ATLAS image reduction pipeline \citep{Tonry_2018_2} subtracts a static-sky ``template'' image from each reduced image.  The template image is created from thousands of historical ATLAS observations and represents the non-varying sky. When the static sky is subtracted, what remains are transient phenomena such as variable stars, supernovae, and most importantly for ATLAS, moving objects --- asteroids and comets.

\begin{figure*}
  \includegraphics[width=\textwidth]{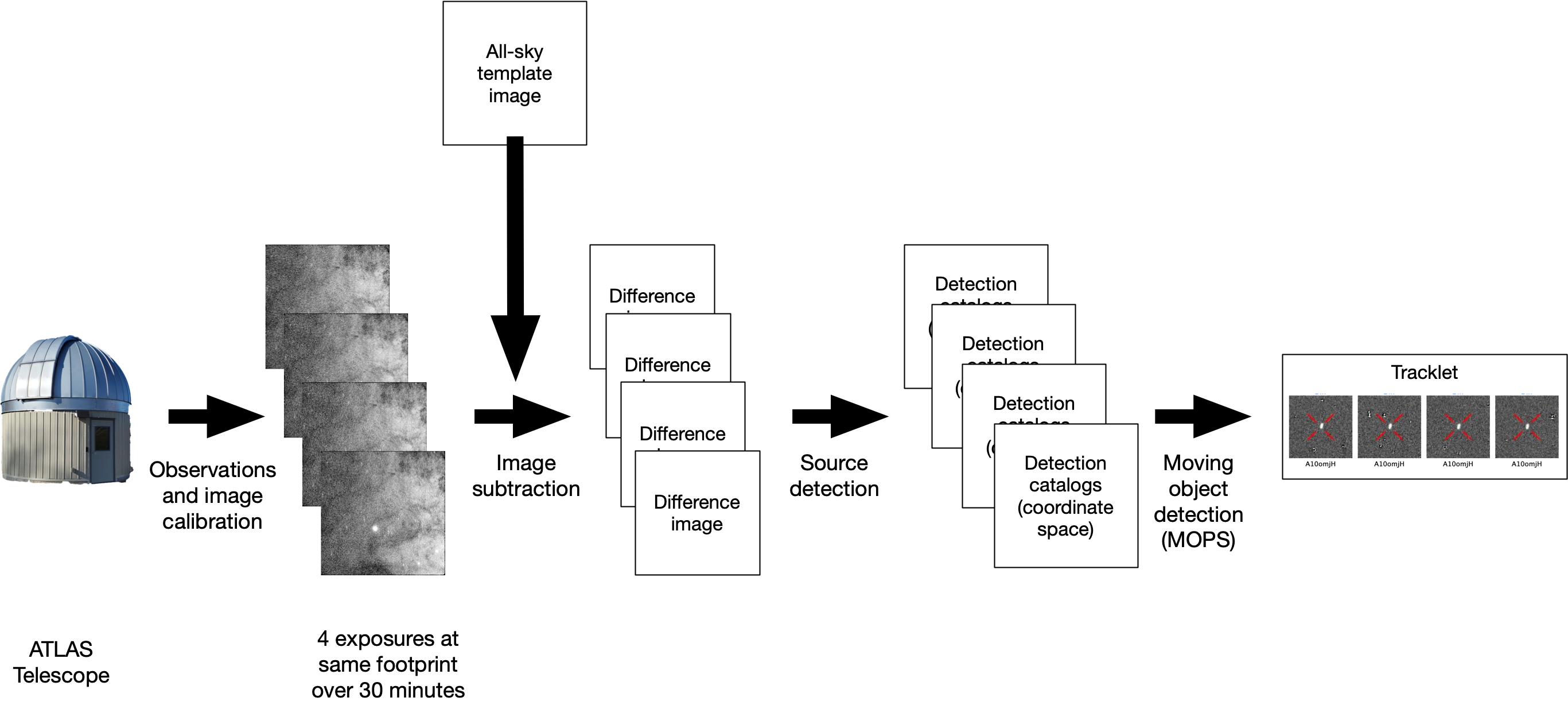}
  \captionof{figure}{\small High-level diagram of ATLAS processing. For each image, the reference sky is registered and subtracted from the source image, leaving moving objects and other transient features. Then the pipeline identifies sources in each subtracted image and saves a catalog of candidate moving object locations. MOPS then links sources from exposure catalogs together to form tracklets. Image subtraction allows ATLAS to detect asteroids very close to bright stars and in the galactic plane. }
  \label{atlas flowchart}
\end{figure*}

After subtraction of the static sky template image, the ATLAS pipeline searches the subtracted image for astronomical sources with signal-to-noise ratio $>5$ and stores their coordinates and other metadata in per-exposure catalogs. The ATLAS moving object processing system (MOPS) \citep{Denneau_2013} examines the catalogs of sources in the subtracted images and creates groupings of detections consistent with linear motion across the sky. A composition of four detections by MOPS, assumed to be a real moving object, then makes a tracklet, the fundamental information block used by ATLAS to report asteroid observations. A visualization of the ATLAS processing pipeline can be found in Figure \ref{atlas flowchart}.

Our model classifies tracklets using small $100\times100$ ``postage stamp'' images of source detections from subtracted images. The MOPS pipeline works in celestial coordinate space and normally does not examine image data due to the computational costs of retrieving pixels for the $\sim10^7$ nightly detections. To integrate our model, we retrieve postage stamps for all tracklets that have been created; this produces a much smaller number ($\sim10^4$) of postage stamps that must be retrieved and classified since most detections in an image do not form a tracklet. Most asteroids are faint and span a small amount of the 16-bit range of allowed pixel values, and therefore the postage stamps are image-equalized for maximum contrast so that relevant features are amplified for the classifier.

Examples tracklets produced by MOPS can be seen in Figure \ref{tracklets}.

\vspace{0.2cm}
{\centering
\includegraphics[scale=0.4]{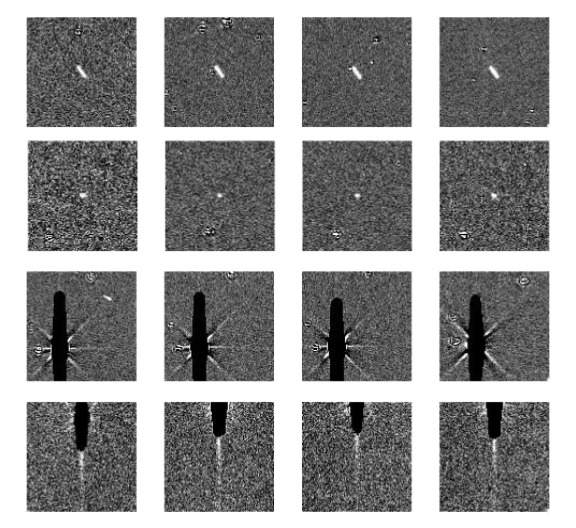}
\captionof{figure}{\small The first row shows streaked NEO 2020 NK1 representing a real tracklet. The second row shows another real tracklet containing asteroid (4700) Carusi. The third and fourth rows are false tracklets caused by bright pixels from optical diffraction spikes and readout optical artifacts respectively. Black pixels represent saturated pixels caused by bright objects.}
\label{tracklets}
\par}

\section{Methodology}

\subsection{Data}


Real astronomical objects are either a solar system object, a variable star, or a point source that is indistinguishable from a slow moving object. Tracklets can be categorized as {\it real} or {\it bogus} depending on whether or not the tracklet corresponds to a moving real astronomical object in the solar system. The bogus category can be divided into sub-categories which represent the majority of sources that cause a bogus tracklet. ATLAS produces its own metrics for every detection in a subtracted image;  these metrics are used to cull the set of detections into probable real astronomical objects. This culling process still leaks many false detections to the MOPS asteroid processing which then can turn into bogus tracklets.


To achieve the highest classification accuracy with a neural network, we used data spanning a wide variety of image types that are captured by the asteroid processing system every night. Types of images that contribute to a real or bogus category, and used to train the neural network, are called {\it classes}, shown in Figure \ref{classes} and in Table \ref{class_desciptions}. There are additional classes that contribute to bogus tracklets, such as optical ghosts (reflected bright sources) and glints (reflections from internal optical elements), or effects from clouds; these have been excluded due to their rarity and difficulty of establishing training data based on their non-uniform appearances.

Table \ref{data sets} summarizes each of the data sets used to train and test our two step deep learning detection classifier.

\vspace{0.2cm}
\begin{center}
  \begin{tabular}{lsss}
    \toprule \toprule
    \multirow{1}{*}{Data Set} &
      {Images} & {Tracklets} & {Training}\\
      \midrule
    Curated & 3,500 & No & Yes\\
    Evaluation & 250,000 & Yes & Yes\\
    \bottomrule
  \end{tabular}
  \captionof{table}{\small Two different data sets consisting of ATLAS images were used to train and evaluate the network. Both data sets were used for training, however, the main goal of the evaluation data set was to test the model's robustness and consistency. The curated data set did not have access to each postage stamp's parent tracklet label as listed in the Tracklets column. Therefore the curated data set cannot be used to train a network to classify tracklets.}
  \label{data sets}
\end{center}

\subsubsection{Curated Data Set} 
\begin{figure*}
  \includegraphics[width=\textwidth]{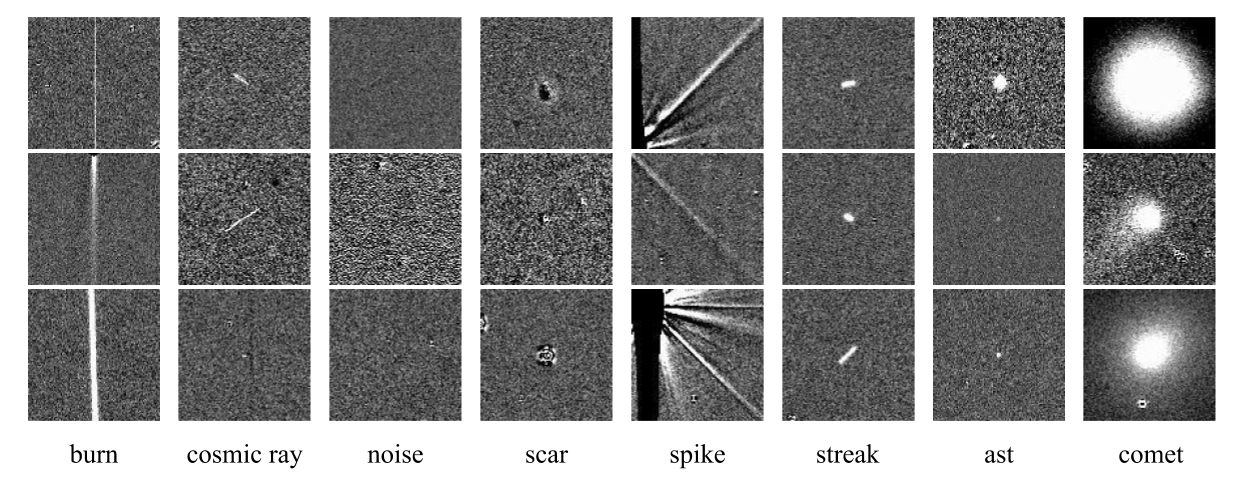}
  \captionof{figure}{\small After  considerable  historical  analysis  of  ATLAS detections, seven classes were chosen to represent all possible types of postage stamps in the curated data set. Each class has unique visual cues but some variations make it indistinguishable from another class. For example, some faint AST can be impossible to distinguish from background noise without additional information.}
  \label{classes}
\end{figure*}

The ATLAS database containing images from real and bogus asteroid tracklets was used to generate a curated data set for neural network training. Images were selected based on an internal non-machine learning software called {\tt vartest05} \citep{Heinze_vartest} that pre-classifies detected sources as one of the classes from Figure \ref{classes}. Even though large amounts of bogus tracklets are parsed everyday, the {\tt vartest05} classifications are not preserved in the processing and therefore it is not trivial to separate detections into their subclasses for training. Additionally, the various subclasses do not occur equally --- STREAK is much less common than SPIKE -- so building a clean data set for these classes with uniform representation  has resulted in a possibly smaller data set than ideal (see work in \citep{Duev_2019} for a CNN trained on a similar task with over 35,000 images).

The complete curated data set consisted of 3500 images each associated with one of seven classes. Furthermore, we balanced the data set so that 500 images were associated with each class. The training set consisted of 470 of the 500 images associated with each class while the rest of the images were used for validation.


\vspace{0.2cm}

\begin{table*}
  \centering
  \begin{tabular}{p{2cm}p{1.2cm}p{1.4cm}p{6cm}p{6cm}}
  \toprule \toprule\\
    \multirow{1}{*}{\textbf{Class}} &
      {\textbf{label}} & {\textbf{Category}} & {\textbf{Description}} & {\textbf{Appearance}}\\\\
      \midrule\\
    Burn & BURN & bogus & vertical electronic readout artifacts caused by bright sources on the detector. & long vertical lines.\\\\ 
    Cosmic Ray & CR & bogus & bright spots in the charged-coupled device (CCD) in the ATLAS camera due to electrons released in the detector silicon by a collision with a high-energy particle. CR artifacts are often very sharp because they only affect a region on the CCD much smaller than a point source imaged through the optical system. & small, sharp, and often non-circular shapes. \\\\
    Noise & NOISE & bogus & detected as a source due to Poisson and readout noise distributions. & no distinguishable object at the position of detection (center of the image) compared to the area around it. \\\\
    Scar & SCAR & bogus & residual bright pixels left over from an imperfect subtraction of a bright star from the sky template image. & clearly identifiable spots with high contrasting color. \\\\
    Diffraction Spike & SPIKE & bogus & caused by bright stars diffracted through the optical path. & dispersing lines appearing near extremely bright sources at predetermined angles (usually 45 degrees). \\\\
    Astronomical Object & AST & real & a real astronomical object; mainly asteroids, variable stars, and satellites but can also contain comets. & brighter than NOISE but usually fainter than CRs and circularly shaped. \\\\
    Streak & STREAK & real & real objects (asteroids or artificial satellites), that leave a linear trail on the image due to its motion during a $\approx 30$ second exposure. & stretch an asteroid sized footprint across part of the image. \\\\
    Bright Comet & COMET & real & comets much wider than the comets in the AST class. & take up a large portion of the image due to their brightness with a shape similar to AST and have some kind of faint trail. \\\\
    \bottomrule
  \end{tabular}
  \captionof{table}{\small Each class has a corresponding Label and Category which represents what kind of tracklet, real or bogus, each class contributes to. A short description of each class's origin and visual appearance in postage stamps can be found in columns 4 and 5.}
  \label{class_desciptions}
\end{table*}

The goal of this data set is to identify the class causing the ATLAS detection rather than the largest or most commonly appearing class. Images containing multiple different classes were manually removed from the training data to see if the network could learn the unique features of each class.

We separated the curated data set into a training set (94\% of the data set) and a validation set (6\% of the data set). The order of each set was randomly shuffled and due to the small size of this curated data set, some minor data augmentation was used to increase class generalization. However, since object shape, sharpness, and magnitude are such a key part of class separation, image augmentations such as random blurring, resizing, and brightness jitters were not used. Instead we used random horizontal/vertical image flipping, rotations, and cropping.

\subsubsection{Evaluation Data Set} Multiple ATLAS completed nights form an evaluation data set of real known,  real unknown, and bogus tracklets. Known tracklets contain a pre-identified object that ATLAS has been following with Inter-Night linking \citep{Denneau_2013}. Unknown tracklets are made of objects that ATLAS cannot match with a known object. Real unknown and bogus tracklets were labeled by human experts while the known tracklets are classified by Inter-night linking. Known tracklets do not require human confirmation but they are still useful as a reference for comparing the performance of the non-machine learning pipeline to the deep learning approach. 

Even though real and known tracklets share similar features, known tracklets have been successfully classified by the current non-machine-learning ATLAS pipeline while unknowns have not. However, they are not necessarily visually different since known tracklets are identified by more than just their postage stamp (tracked from previous nights based on position, size, rate of motion).

This data set consisted of two lunations of ATLAS observations from May 7-July 5 2020. The first lunation was used for training, and the second for validation. The training and validation data consisted of pre-labeled real and bogus tracklets. The real tracklets were further labeled as ``known'', meaning that a tracklet matched the position of an asteroid in the MPC catalog, and ``unknown'', meaning that the tracklet could not be matched to a catalogued object at the time of detection. ATLAS detects about 1000 times more known objects than unknown objects. The orbits for known objects are accurate enough that tracklets can be automatically assigned to a known object. 

Each lunation of unknowns consisted of $\approx 27,600$ tracklets (over 110,400 images), approximately 98\% of which are bogus tracklets and 2\% are real unknown tracklets. On the other hand, each lunation of knowns consisted of $\approx 200,000$ tracklets (over 800,000 images) all of which are real known tracklets. No image augmentations were applied to the tracklet postage stamps during training due to the larger size of the data set.

\begin{figure*}
  \includegraphics[width=\textwidth]{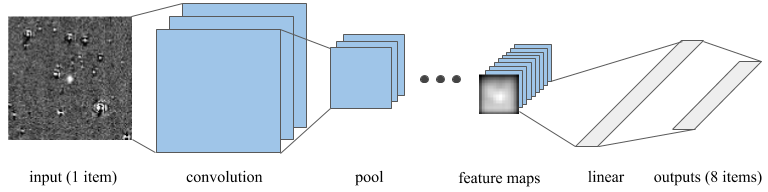}
  \captionof{figure}{\small In a convolutional neural network such as resnet-18, filters are applied to the original image (convolution layer) and resized (pool layer) multiple times. The feature maps produced by the final pooling layer are flattened into a 1 dimensional tensor and passed into a dense layer (linear layer) which results in an output for each class. The output for each class is a singular number between zero and one which represents a confidence measurement of the detection being that class (the sum of all the output values is equal to one).}
  \label{ICN}
\end{figure*}

\subsection{Network Architecture}

Since each tracklet can contain detections of different classes, we designed a two-step model consisting of an Image Classification Network (ICN) and Tracklet Classification Network (TCN) to tackle image and tracklet classifications respectively. Due to the high visual similarities between some class variations and faint asteroids, see Figure \ref{noisevsfaint}, this two step model was prioritized over a single fully connected neural network. By having the network classify in terms of all eight classes, we can better understand how the network is separating each class and which ones it commonly misclassifies as another class. Unknown tracklets removed before human screening are essentially unobserved by ATLAS and could remain undetected to other asteroid surveys. For this reason, we prioritized the preservation of real tracklets over a lower false positive rate while designing and training the model.

\vspace{0.2cm}
{\centering
\includegraphics[scale=0.7]{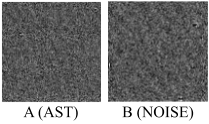}
\captionof{figure}{\small Image A and B show an example of two different classes that are visually indistinguishable from another. Image A shows a faint main belt asteroid (117708) and image B shows a false detection caused by noise. By classifying each postage stamp in a tracklet with eight different classes, the network can associate error margins for each class based on the result of all the other classes.}
\label{noisevsfaint}
\par}
\vspace{0.2cm}

The PyTorch library \citep{NEURIPS2019_9015} was used to generate the neural networks and training was done with a single CUDA enabled GPU to decrease training times. Each network's training parameters are summarized in Table \ref{model parameters}.

\vspace{0.2cm}
\begin{center}
  \begin{tabular}{lll}
    \toprule \toprule\\
    \multirow{1}{*}{Parameters} &
      {ICN} & {TCN}\\
      \midrule
    Optimizer & Adam & SGD\\
    Criterion & CE & MSE\\
    Learning Rate & 0.01 & 0.1\\
    Data Set & Curated & Evaluation\\
    Batch Size & 15 & 1\\
    \bottomrule
  \end{tabular}
  \captionof{table}{\small The data set row represents the inputs used to train the model and the criterions or loss column used either Cross Entropy Loss (CE) or Mean Squared Error Loss (MSE). A decaying learning rate was used alongside early stopping to reduce over-fitting and training runtime.}
  \label{model parameters}
\end{center}

\subsubsection{Image Classification Network} For the image classifier, we created a convolutional neural network (CNN) that takes an individual image and returns a set of eight confidence scores, one for each class. At first, we attempted to train a custom CNN but simple tests with the curated data set showed limited performance when compared to a pre-trained resnet-18 network \citep{he2015deep}. It is likely that the pre-trained filters and convolutional layers (Figure \ref{ICN}) on ImageNet contributed to the higher class accuracy \citep{russakovsky2014imagenet}. We fine-tuned the pre-trained resnet-18 by training it on the curated data set.


It is important that the classification runtime be kept as low as possible so that we can minimize the latency for processing time. Therefore, we decided to employ a shallower network with less layers over a deeper network to help decrease computation time. Also, we theorized that the seemingly low visual complexity of the classes and small number of classes would not be able to reap the benefits of deeper networks. One reason for this assumption is a deep network's likelihood to over-fit much faster during training than a simple neural network. This approach has been used with success in recent research similar to this one \citep{Duev_2019}. Furthermore, training on the curated data set with multiple deep network architectures such as vgg-19, resnet-101, densenet-121 among others showed no accuracy gains \citep{simonyan2014deep} \citep{he2015deep} \citep{huang2016densely}.

\begin{figure*}
  \centering
  \includegraphics[scale=0.4]{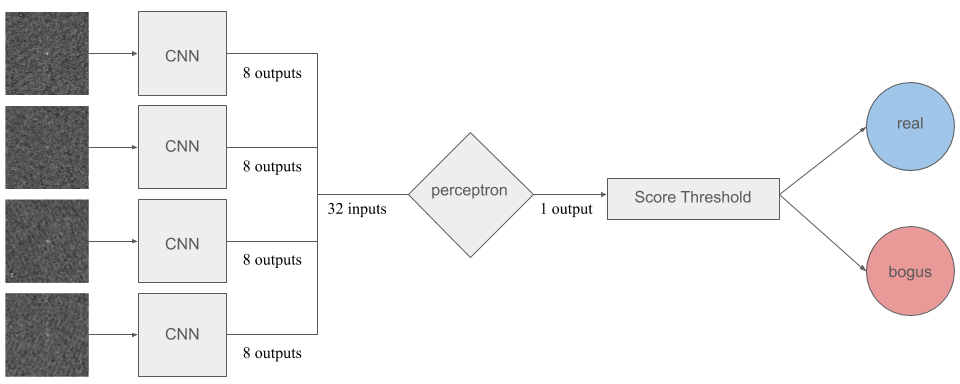}
  \captionof{figure}{\small Four postage stamps from a tracklet are passed into a CNN one by one and a total of 32 confidence scores are generated. These 32 values are passed into the Multi-Layered Perceptron (MLP) in the same order for each tracklet so that the network can learn the optimal weights for each input. A single value between zero and one is produced by the MLP where 1 represents a real tracklet and 0 represents a bogus tracklet. This type of model structure allows us to train an MLP that takes in more or less than 4 postage stamps for uncommon tracklets as long as a training data set is provided.}
  \label{TCN}
\end{figure*}

\subsubsection{Tracklet Classification Network} To classify tracklets, we used a multi-layered perceptron (MLP) that takes the outputs of the ICN as inputs. MLPs are well known among feed forward neural network architectures and represent one of the simplest deep learning architectures commonly used. By stacking deeper layers of perceptrons (takes multiple inputs and spits out one output) more complex details and features can be extracted from the input. The MLP returns a new set of confidence scores for the tracklets as bogus or real, see Figure \ref{TCN}. Since an MLP takes an input of fixed size only the class confidence scores of the first four postage stamps in a tracklet are provided to the TCN as inputs. Approximately 90\% of tracklets contain at least four detections.

The MLP cannot be trained with the curated data set since the postage stamps in the data set are not linked to a tracklet label. Instead, we used the evaluation data set which only provided real and bogus as ground truth labels for each tracklet. The use of the evaluation data set for training will remove the MLP's potential to classify tracklets as a class rather than a category.

The confidence scores of some classes for a postage stamp will be more competitive than others (when the standard deviation of all the confidence scores is small). This is simply due to the fact that some class cases are practically indistinguishable from another class and an image can contain several different classes. Additionally, we expect most postage stamps in a tracklet to return similar confidence outputs when passed through the ICN. However, it is possible for tracklets to contain radically different postage stamps with unique features respective to another. This could cause different confidence outputs for postage stamps in the same tracklet. Therefore, the goal of the TCN is to determine a set of rules with the confidence scores that would maximize correct tracklet classification.

To ensure a low false negative rate, we designed the neural network with a real tracklet bias using a simple weighted loss scheme. By assigning weights ($\alpha_1,\alpha_2$) to both the real and bogus category in the MSE Loss function ($\Tilde{L}$) which takes in the networks predictions ($x$) and the ground truth ($y$), we can generate an imbalance/bias for real tracklets.
\begin{align}
    \alpha(y)=\left\{\begin{array}{cc}
    \alpha_1\;{\rm if}\;y\;{\rm is\;bogus}\\
    \alpha_2\;{\rm if}\;y\;{\rm is\;real}
    \end{array}\right.\\
    L(x,y) = \alpha (y) \Tilde{L}(x, y)
    \label{weighted loss}
\end{align}
with $\alpha_1<\alpha_2$ and where $L$ is the weighted loss function. 
These adjustments come at the cost of a lower accuracy on bogus tracklet classification. Multiple tests were required to identify an optimal balance in minimizing both the false negative and false positive rate. We found that $\alpha_1=1$ and $\alpha_2=4$ provided the best results for increasing real tracklet classification accuracy.

\section{Results}

\begin{figure*}
\centering
\includegraphics[scale=0.37]{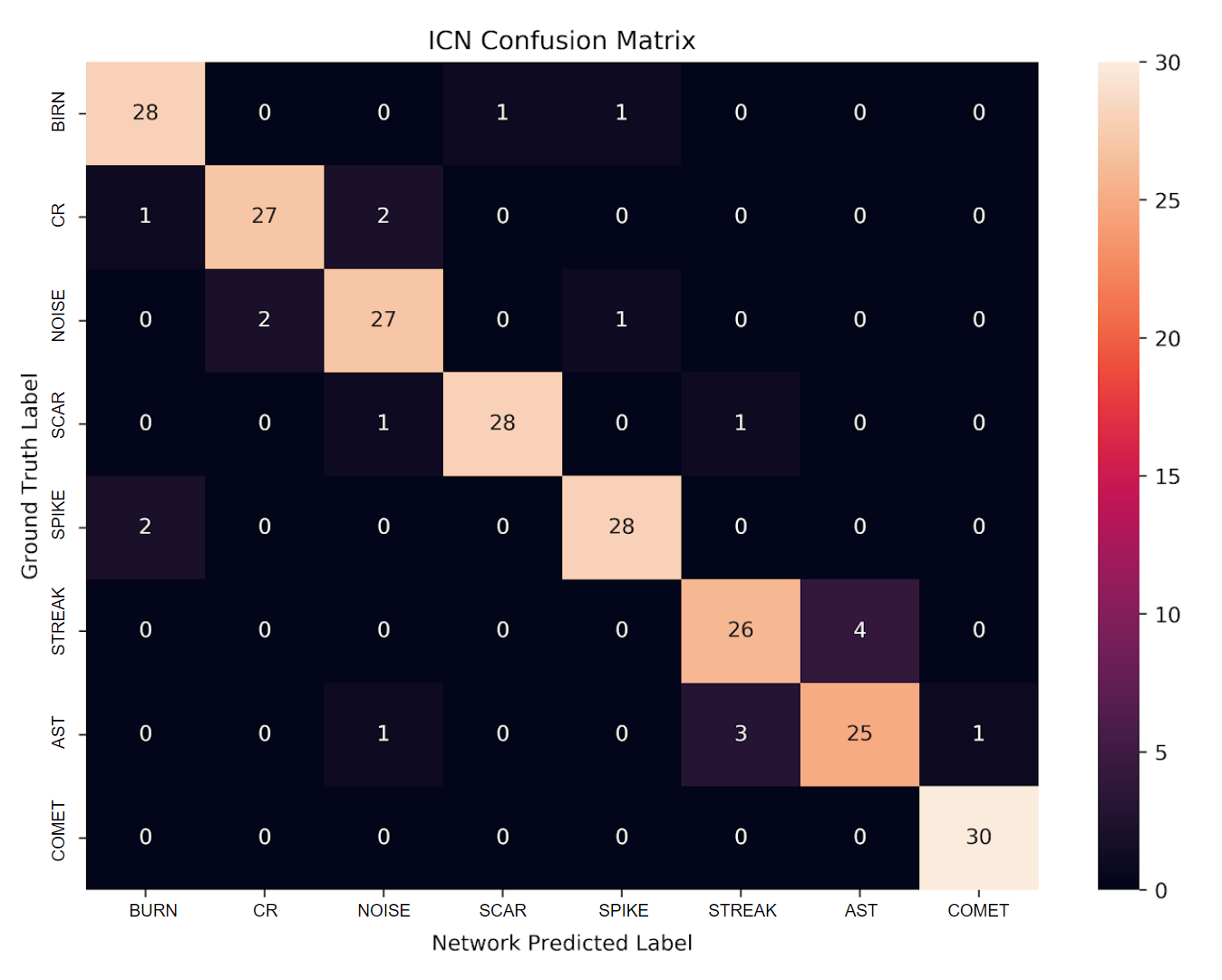}
\captionof{figure}{\small Each class was tested on 30 standalone validation images from the curated data set. A confusion matrix using the validation images was generated at every epoch. The matrix seen above corresponds to the last epoch used to train the official model.}
\label{confusion matrix 7}
\end{figure*}

We will first address the capability of the ICN by generating a confusion matrix based on its performance on the validation images in the curated data set, see Figure \ref{confusion matrix 7}. It can be observed that the ICN does best separation for SCAR and SPIKEs. The primary issue of our ICN, highlighted by the confusion matrix, is that about 13\% of images in the STREAK class are misclassified as AST. Given that STREAKs consists of NEOs and small STREAKs are barely distinguishable from comets and other asteroids, it is expected that the ICN has difficulty differentiating between the two. However, since both STREAK and AST classes represent image types from a real tracklet it should not affect the overall accuracy of the model.

We trained the TCN on a month of ATLAS tracklets from May 5 to June 4 2020. From Figure \ref{confusion matrix 2} it can be deduced that about 99.6\% of real tracklets were correctly classified, while 90.8\% of bogus tracklets were correctly classified. The bogus tracklet accuracy is lower than the real tracklet accuracy due to the weighted loss scheme highlighted in section 3.2.2. Additionally, the ICN was trained on the curated data set which contained more distinguishable artifacts and image types than the images in the evaluation data set. This issue is not as impactful for postage stamps from real tracklets since the evaluation data set is heavily represented by tracklets the ATLAS detection pipeline classified as real and are therefore similar to the AST, STREAK, and COMET type images in the curated data sets. Also, there is generally more variation in the postage stamps from bogus tracklets than the postage stamps from real tracklets containing asteroids and comets. All of these effects result in a lower bogus tracklet accuracy when compared to the real tracklets.

{\centering
\includegraphics[scale=0.4]{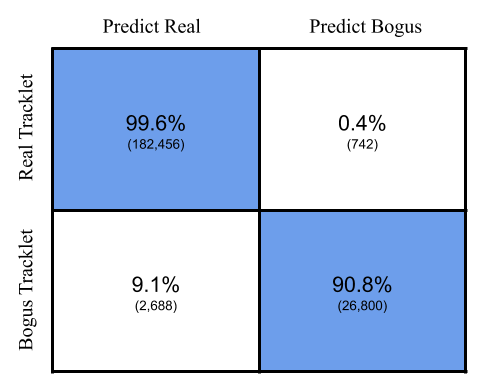} 
\captionof{figure}{\small The TCN was tested on 30 nights of ATLAS data from June 5 to July 5 2020. A total of 212,686 tracklets were classified from the evaluation data set. Our priority was to minimize the false-negative rate which are tracklets the model deems bogus but are in reality real NEO.}
\label{confusion matrix 2}
\par}
\vspace{0.2cm}

The Receiver Operating Characteristic (ROC) shown in Figure \ref{ROC} helps us understand the cost of a higher true positive rate versus an increase in false positive rate. The ROC curve shows that for a threshold of 0.15 false positive rate we can reach a 0.97-0.99 true positive rate. The ROC curve also shows that optimizing the false positive threshold above 0.1 gives diminishing returns for the true positive rate. On the other hand, a threshold of 0.02 or smaller will drastically reduce the true positive rate of the model which is a higher priority to ATLAS than a slightly lower false positive rate. Finally, the 91\% fraction of correctly predicted bogus tracklets leads to significant reduction in human workload of tracklet review, at a cost of 0.4\% incorrectly classified real objects.

{\centering
\includegraphics[scale=0.5]{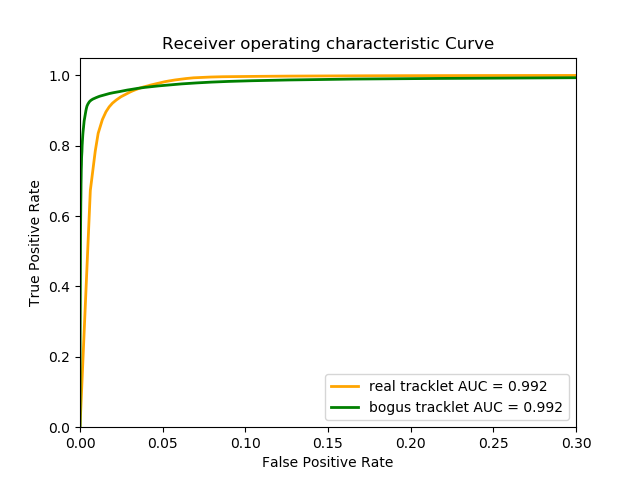} 
\captionof{figure}{\small The Receiver Operating Characteristic (ROC) curve of the TCN's performance on the evaluation data. When the model clearly distinguishes between true positives and false positives the curve will converge to 1 faster which means the Area Under the Curve (AUC) will approach 1). The AUC for real tracklets is 0.992 which indicates that the model does a good job at separating real tracklets from bogus tracklets. The AUC for bogus tracklets is equivalent to the AUC for real tracklets because it is a binary classification problem.}
\label{ROC}
\par}
\vspace{0.2cm}



The high accuracy of the trained ICN and TCN was reproduced multiple times on the curated and evaluation data set respectively in case any parts of the model are permanently lost.

\vspace{0.2cm}
\begin{table*}
  \centering
  \begin{tabular}{lcrcrcrcrcrc}
    \toprule \toprule\\
    \multirow{1}{*}{Tracklet} &
      {I} & {Score} & {II} & {Score} & {III} & {Score} & {IV} & {Score} & {Pred} & {Score} & {Truth}\\
      \midrule
    1 & ST & 0.99 & ST & 1.0 & ST & 1.0 & ST & 1.0 & real & 0.904 & real\\
    2 & AS & 0.99 & AS & 0.90 & AS & 0.99 & AS & 0.99 & real & 0.999 & real\\
    3 & SP & 0.66 & SP & 0.80 & SP & 0.71 & SP & 0.94 & bogus & 0.058 & bogus\\
    4 & BU & 0.78 & BU & 0.91 & BU & 0.48 & BU & 0.63 & bogus & 0.129 & bogus\\
    5 & NO & 0.63 & AS & 0.82 & NO & 0.92 & AS & 0.98 & bogus & 0.740 & real\\
    6 & AS & 0.96 & NO & 0.99 & AS & 0.85 & AS & 0.98 & real & 0.982 & bogus\\
    \bottomrule
  \end{tabular}
  \captionof{table}{\small Each row shows the ICN's highest scoring class on each individual image and the TCN's prediction for the tracklet. The output scores for each classification can be found to the right of its label.  Also, the ground truth for each tracklet is shown in the last column. Note that the prediction score differs from the other scores by indicating a bogus tracklet if the score is closer to zero and indicating a real tracklet if the score is closer to one. A threshold tracklet score of 0.8 is used to determine real versus bogus classification (e.g. score of 0.6 is real). The class labels were simplified to BU, CR, NO, SC, SP, ST, AS, CO.}
  \label{qualitative results}
\end{table*}

A qualitative analysis of the TCN's predictions show that although it consistently identifies more prominent artifacts, it has some trouble with subtle cases involving noise as seen in Figure \ref{results tracklets} and Table \ref{qualitative results}. 

\vspace{0.2cm}
{\centering
\includegraphics[scale=0.35]{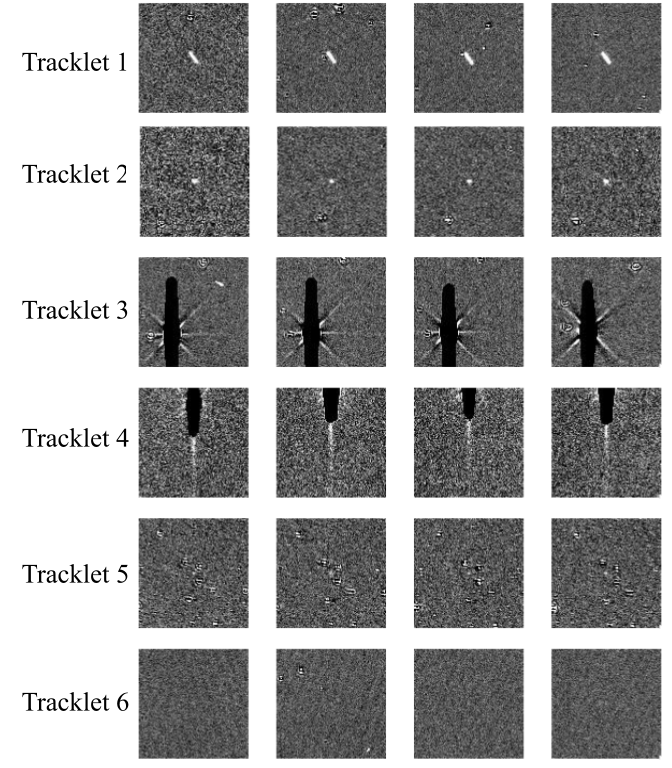}
\captionof{figure}{Several example tracklets are shown above and their corresponding model predictions are listed in Table \ref{qualitative results}. The first four tracklets are the same as the ones from Figure \ref{tracklets}. Tracklet 5 is main belt asteroid (17710) and Tracklet 6 consists of false detections caused by noise. Both of these tracklets were handpicked from the evaluation data set to show example tracklets that the deep learning model struggles to correctly classify.}
\label{results tracklets}
\par}
\vspace{0.2cm}

In the unknown evaluation data set results, the two-step model never failed to classify tracklets such as tracklet 1 and 2 as real. Additionally, the model consistently classified prominent SPIKEs and BURNs as displayed in tracklet 3 and 4. Most of the false negative results came from the ICN's inability to consistently differentiate NOISE from AST. In tracklet 5 an extremely faint astronomical object is positioned at the center of each image while surrounded by subtraction artifacts (SCAR type artifacts). The ICN classifies most of the images as NOISE with high confidence scores which results in the TCN classifying the tracklet as bogus. In tracklet 6 the ICN classifies most of the postage stamps as AST even though the detections are indistinguishable from Poisson noise.

The two-stage deep learning model is currently deployed on ATLAS as a primary filter prior to NEO screening before submission to the MPC. The model is currently undergoing preliminary inspection as it is compared to the NEO candidate list generated by ATLAS before the deep learning removes bogus tracklets. Several weeks of deployment have shown encouraging results as it reduces the NEO candidate list by $\approx$ 95\% without losing any real tracklets.

Table \ref{deployement results} shows the real-world performance of the TCN against unknown asteroid candidates that must be reviewed prior to submission to the MPC.  The false positives (column FP) are dominated by a single type of artifact (a bright horizontal optical effect caused by visible planets close to the field) that was unknown to the ICN for this work.  Training the ICN on this additional image type would effectively remove this type of false tracklet, resulting in FP rates near 1\% or less.

Comparison of the TCN with human screening is somewhat delicate.  Because of the comprehensive job the human must perform, the TCN  cannot screen as accurately as its human counterpart.  By definition, human-screened tracklets are 100\% correct since they form the basis for the training sets.  But for visual screening only, we find that the TCN performs very accurately for image types where the ICN has been trained, 

The human screening process also evaluates criteria beyond what the TCN was designed for.  The TCN filters tracklets purely based on visual appearance, {\it i.e.} imagery only, while humans screen both on visual appearance and on dynamical parameters of a tracklet. Typical scenarios are a tracklet composed of detections from multiple variable stars, or two different real asteroids mis-linked into a single tracklet. In these cases a ``fit'' to the motion may produce plausible values, but further inspection of positional residuals against the fitted motion would show that at least one of the sources shows no actual motion and is consistent with a variable star.  To a human, this is a bogus but ``real-looking'' tracklet, while the TCN simply (and correctly) considers this ``good'', since the images show detections that are indistinguishable from asteroids.

A natural extension of our model would be the inclusion of these additional non-image parameters, or metadata, into an additional machine-learning network that complements the ICN and TCN.  ATLAS provides numerous parameters about every detection that would be suitable: X, Y location on the detector, proximity to bright stars and planets, proximity to known electronic artifacts, and so on. The flexibility of the model could allow us to feed image specific metadata along the images in the ICN while tracklet specific metadata would be fed along the outputs of the ICN into the TCN. However, we have chosen not to explore this avenue in this work for two reasons: a) the number of available metadata parameters is large (on the order of dozens) and the effort to train and understand a metadata model would distract from the core of this work; and b) ATLAS already performs non-machine-learning classifications of detections with the metadata parameters using a code called {\tt vartest05}, and it is our opinion that {\tt vartest05} is effective enough at pre-classifying detections so that the available performance gains are minimal.  {\tt vartest05} employs its own internal model of how a bright source can create a burn or diffraction spike, or what a cosmic ray might look like, and provides a score for likelihood of being one of its known detection types. The set of detections that is seen by a human reviewer is thus pre-screened by {\tt vartest05}; this screened set still contains many false detections, and that is our motivation for the ICN and TCN created for this work.


\section{Discussion}

\vspace{0.2cm}
\begin{table*}
  \centering
  \begin{tabular}{lrrrrr}
    \toprule \toprule\\
    \multirow{1}{*}{Night} & {Unknown} & {Real} & {FP (\%)} & {FN (\%)}\\
      \toprule
    59090 & 188 & 26 & 6.4 & 0 \\
    \midrule
    59091 & 11 & 7 & 9.1 & 0\\
    \midrule
    59092 & 79 & 8 & 1.3 & 0\\
    \midrule
    59093 & 637 & 8 & 4.1 & 0\\
    \midrule
    59094 & 185 & 11 & 3.3 & 0\\
    \midrule
    59095 & 971 & 22 & 4.1 & 0\\
    \midrule
    59096 & 47 & 16 & 2.1 & 0\\
    \midrule
    59097 & 413 & 12 & 0.2 & 0\\
    \midrule
    59098 & 29 & 16 & 3.4 & 0\\
    \midrule
    59099 & 67 & 25 & 1.5 & 0\\
    \midrule
    59100 & 134 & 63 & 2.2 & 0\\
    \midrule
    59101 & 314 & 75 & 2.5 & 0\\
    \bottomrule
  \end{tabular}
  \captionof{table}{\small The deployed model's results were recorded from August 29, 2020 (MJD 59090) to September 9, 2020 (MJD 59101) using production ATLAS data to assess its performance. ``Unknown'' is the number of tracklets that could not be automatically matched with a known object, and ``Real'' is the number of these that were real astronomical objects based on visual inspection.  ``FP'' is the false positive rate (bogus scored as real) and ``FN'' the number of real scored as bogus. The variation in the number of tracklets each night is due to sky coverage, weather conditions, and the presence of sky features that can produce false tracklets (bright stars, planets). The false negative rate over this interval was zero, meaning no real tracklets were lost due to misclassification.}
  \label{deployement results}
\end{table*}

While a real-bogus threshold of $0.5$ is a natural and sensible threshold to separate real and bogus tracklets, practical considerations related to the endeavor of NEO discovery lead to asymmetric priorities that can inform the selection of a rejection threshold.  Unknown NEOs detected by ATLAS are by definition possibly dangerous to Earth, and therefore the loss of even a single NEO due to a false negative assessment can have major consequences.  Asteroid 2019~OK, a dangerous 100 m diameter NEA discovered by the SONEAR survey in 2019, passed within 70,000 km of the Earth one day later \citep{2019_OK}.  2019~OK was imaged three days prior to the SONEAR observations by ATLAS, but the tracklet was moving very slowly on the sky, resembling a stationary object, and the tracklet did not pass other quality cuts in the system so it was not reported to the MPC immediately and three days of warning time was lost.

Conversely, the costs of a false positive assessment by the system are a) the expenditure of costly human and telescope resources to chase nonexistent objects, and b) the contamination of the discovery stream of data from bogus objects and gradual reduction in confidence in the NEO discovery system. For ATLAS, human review of tracklets prior to submission is the backstop against submission of bogus tracklets to the MPC. Preservation of detected NEOs outweighs other considerations, and therefore in this work we have biased our model toward keeping as many real tracklets as possible instead of biasing toward minimization of false positives. Adjustments to the weighted loss function and training data for both the ICN and TCN are required to adapt the model to different priorities.

We selected a confidence score threshold of 0.8 on the TCN's output to discriminate real from bogus tracklets. This threshold was chosen based on the model's results on the evaluation data set. A more quantitative analysis of the testing may provide insight towards a new threshold that would decrease the false positive rate without increasing the false negative rate. Theoretically, a lower threshold should allow more real tracklets to be correctly classified but will allow more bogus tracklets to be incorrectly classified. Similarly, if the threshold is increased more bogus tracklets will be correctly classified while more real tracklets will be incorrectly classified.

During our deployment, we found that bright comets were often incorrectly classified as burns (BU). To address this issue we created a curated comet class, which removed the problem. Another common type of misclassification came from asteroids detected near the edge of the CCD. The postage stamp images for these detections have a horizontal or vertical linear feature (due to the CCD edge) that looks like a burn to the classifier. We have not yet retrained the model against these edge artifacts, but they are detectable in downstream processing that has access to the CCD coordinates of a tracklet.

Among the evaluation data set there are false tracklets composed of variable stars whose brightness has increased against their average brightness and therefore appear as a new astronomical source in an ATLAS exposure. Even though the detections are real astronomical sources (stars), these tracklets are labeled as bogus because they are not asteroids. Variable stars are visually indistinguishable from an asteroid if images are classified one at a time (e.g. step one of our two step model) because most asteroids resemble a star in a single 30-second ATLAS exposure, but the model (correctly) classifies them real. A random sampling of our input training set suggests that the bogus tracklets from the evaluation data set consists of 5-10\% variable stars. Ignoring these would result in a much higher false positive rate than shown in the results section.

Finally, we recognize that the curated data set is small compared to some data sets used to train CNNs, such as ImageNet \citep{NIPS2012_4824} \citep{russakovsky2014imagenet}. ImageNet, a large data set containing hundreds of classes and millions of images, has been used to train several state of the art neural networks due to the quantity of training data it presents. It is likely that a larger training pool would result in better performance decreasing the false positive rate while keeping the true positive rate equal or above the current response.

\section{Conclusion}

We have designed and deployed a lightweight machine classification model that can accurately discriminate between real and bogus tracklets in the ATLAS asteroid detection pipeline.  This model achieves a $\sim$90\% reduction in the workload of false tracklets to review each night at a cost of 0.4\% in real objects.  This improved accuracy is an essential step toward immediate, automatic submission of dangerous asteroids to the MPC after they are detected by ATLAS.  The reduced latency between detection and reporting increases the ability of follow-up telescopes to track inbound asteroids because their positions will be closer to their discovery positions and therefore easier to observe.  Perhaps more importantly, reduced latency provides greater warning time for civil defense purposes in the case of an actual impact.  

The model's performance is still being monitored with the goal of identifying specific areas (types of incoming ATLAS data) in which the deep learning model fails to be advantageous over full human screening. We identified eight dominant image types that appear in ATLAS tracklets, and unsurprisingly the model performs poorly classifying tracklets with image types that the ICN was not trained on. Since the model was deployed, we have identified and started creating training sets for new classes to improve the model's accuracy. One of these new classes will address horizontal line artifacts generated by very-bright astronomical sources such as the visible planets. Early training results on these bright horizontal features show that they can be effectively removed completely, but this capability was not integrated in time for this work.  Future user feedback on the screening lists will allow us to iteratively improve the accuracy and overall robustness of the model.

\section{Acknowledgements}

This work has made use of data from the Asteroid Terrestrial-impact Last Alert System (ATLAS) project. LD and ATLAS are primarily funded by NASA grants NN12AR55G, 80NSSC18K0284, and 80NSSC18K1575; byproducts of the ATLAS NEO search include images and catalogs from the survey area. The ATLAS science products have been made possible through the contributions of the University of Hawaii Institute for Astronomy, the Queen's University Belfast, the Space Telescope Science Institute, and the South African Astronomical Observatory. This work was supported in part by a National Science Foundation Research Experience for Undergraduate grant (6104374) to the Institute for Astronomy at the University of Hawaii-Manoa. We would like to thank Dr. Michael Bottom, Dr. Robert Jedicke, and Dr. Ben Shappee for their insightful comments and feedback.

\end{multicols}

\newpage

\bibliographystyle{plainnat}
\bibliography{references}

\begin{thebibliography}{21}
\providecommand{\natexlab}[1]{#1}
\providecommand{\url}[1]{\texttt{#1}}
\expandafter\ifx\csname urlstyle\endcsname\relax
  \providecommand{\doi}[1]{doi: #1}\else
  \providecommand{\doi}{doi: \begingroup \urlstyle{rm}\Url}\fi

\bibitem[A. et~al.(2020, in prep)A., L., and J.L]{Heinze_vartest}
Heinze A., Denneau L., and Tonry J.L.
\newblock Algorithms for real/bogus filtering of atlas asteroid and transient
  detections, 2020, in prep.

\bibitem[Alvarez et~al.(1980)Alvarez, Alvarez, Asaro, Michel, Alvarez, Alvarez,
  Asaro, and Michel]{Alvarez80extraterrestrialcause}
Luis~W. Alvarez, Walter Alvarez, Frank Asaro, Helen~V. Michel, Luis~W. Alvarez,
  Walter Alvarez, Frank Asaro, and Helen~V. Michel.
\newblock Extraterrestrial cause for the cretaceous-tertiary extinction.
\newblock \emph{Science}, 208:\penalty0 1095--1108, 1980.

\bibitem[Baron(2019)]{baron2019machine}
Dalya Baron.
\newblock Machine learning in astronomy: a practical overview, 2019.

\bibitem[Center(2019 OK, MPEC 2019-O56)]{2019_OK}
Minor~Planet Center.
\newblock \url{https://minorplanetcenter.net/mpec/K19/K19O56.html}, 2019 OK,
  MPEC 2019-O56.
\newblock Online; accessed 29 January 2014.

\bibitem[Denneau et~al.(2013)Denneau, Jedicke, Grav, Granvik, Kubica, Milani,
  Vereš, Wainscoat, Chang, Pierfederici, and et~al.]{Denneau_2013}
Larry Denneau, Robert Jedicke, Tommy Grav, Mikael Granvik, Jeremy Kubica,
  Andrea Milani, Peter Vereš, Richard Wainscoat, Daniel Chang, Francesco
  Pierfederici, and et~al.
\newblock The pan-starrs moving object processing system.
\newblock \emph{Publications of the Astronomical Society of the Pacific},
  125\penalty0 (926):\penalty0 357–395, Apr 2013.
\newblock ISSN 1538-3873.
\newblock \doi{10.1086/670337}.
\newblock URL \url{http://dx.doi.org/10.1086/670337}.

\bibitem[Duev et~al.(2019)Duev, Mahabal, Ye, Tirumala, Belicki, Dekany,
  Frederick, Graham, Laher, Masci, Prince, Riddle, Rosnet, and
  Soumagnac]{Duev_2019}
Dmitry~A Duev, Ashish Mahabal, Quanzhi Ye, Kushal Tirumala, Justin Belicki,
  Richard Dekany, Sara Frederick, Matthew~J Graham, Russ~R Laher, Frank~J
  Masci, Thomas~A Prince, Reed Riddle, Philippe Rosnet, and Maayane~T
  Soumagnac.
\newblock Deepstreaks: identifying fast-moving objects in the zwicky transient
  facility data with deep learning.
\newblock \emph{Monthly Notices of the Royal Astronomical Society},
  486\penalty0 (3):\penalty0 4158--4165, 04 2019.
\newblock ISSN 0035-8711.
\newblock \doi{10.1093/mnras/stz1096}.
\newblock URL \url{https://doi.org/10.1093/mnras/stz1096}.

\bibitem[Foschini et~al.(2018)Foschini, Gasperini, Stanghellini, Serra,
  Polonia, and Stanghellini]{foschini2018atmospheric}
L.~Foschini, L.~Gasperini, C.~Stanghellini, R.~Serra, A.~Polonia, and
  G.~Stanghellini.
\newblock The atmospheric fragmentation of the 1908 tunguska cosmic body:
  reconsidering the possibility of a ground impact, 2018.

\bibitem[{Harris} and {D'Abramo}(2015)]{2015Icar..257..302H}
Alan~W. {Harris} and Germano {D'Abramo}.
\newblock {The population of near-Earth asteroids}.
\newblock \emph{icarus}, 257:\penalty0 302--312, September 2015.
\newblock URL \url{https://ui.adsabs.harvard.edu/abs/2015Icar..257..302H}.

\bibitem[He et~al.(2015)He, Zhang, Ren, and Sun]{he2015deep}
Kaiming He, Xiangyu Zhang, Shaoqing Ren, and Jian Sun.
\newblock Deep residual learning for image recognition, 2015.

\bibitem[Heinze et~al.(2020, in prep)Heinze, Denneau, Tonry, Weiland, Stalder,
  Rest, Smith, and Smartt.]{Heinze_AU}
A.~N. Heinze, L.~Denneau, J.~L. Tonry, H.~Weiland, B.~Stalder, A.~Rest, K.~W.
  Smith, and S.~J. Smartt.
\newblock Neo population, velocity bias, and impact risk from an atlas
  analysis, 2020, in prep.

\bibitem[Huang et~al.(2016)Huang, Liu, van~der Maaten, and
  Weinberger]{huang2016densely}
Gao Huang, Zhuang Liu, Laurens van~der Maaten, and Kilian~Q. Weinberger.
\newblock Densely connected convolutional networks, 2016.

\bibitem[Khan et~al.(2020)Khan, Sohail, Zahoora, and Qureshi]{Khan_2020}
Asifullah Khan, Anabia Sohail, Umme Zahoora, and Aqsa~Saeed Qureshi.
\newblock A survey of the recent architectures of deep convolutional neural
  networks.
\newblock \emph{Artificial Intelligence Review}, Apr 2020.
\newblock ISSN 1573-7462.
\newblock \doi{10.1007/s10462-020-09825-6}.
\newblock URL \url{http://dx.doi.org/10.1007/s10462-020-09825-6}.

\bibitem[Krizhevsky et~al.(2012)Krizhevsky, Sutskever, and
  Hinton]{NIPS2012_4824}
Alex Krizhevsky, Ilya Sutskever, and Geoffrey~E Hinton.
\newblock Imagenet classification with deep convolutional neural networks.
\newblock In F.~Pereira, C.~J.~C. Burges, L.~Bottou, and K.~Q. Weinberger,
  editors, \emph{Advances in Neural Information Processing Systems 25}, pages
  1097--1105. Curran Associates, Inc., 2012.
\newblock URL
  \url{http://papers.nips.cc/paper/4824-imagenet-classification-with-deep-convolutional-neural-networks.pdf}.

\bibitem[Lieu et~al.(2019)Lieu, Conversi, Altieri, and Carry]{Lieu_2019}
Maggie Lieu, Luca Conversi, Bruno Altieri, and Benoît Carry.
\newblock Detecting solar system objects with convolutional neural networks.
\newblock \emph{Monthly Notices of the Royal Astronomical Society},
  485\penalty0 (4):\penalty0 5831–5842, Mar 2019.
\newblock ISSN 1365-2966.
\newblock \doi{10.1093/mnras/stz761}.
\newblock URL \url{http://dx.doi.org/10.1093/mnras/stz761}.

\bibitem[Paszke et~al.(2019)Paszke, Gross, Massa, Lerer, Bradbury, Chanan,
  Killeen, Lin, Gimelshein, Antiga, Desmaison, Kopf, Yang, DeVito, Raison,
  Tejani, Chilamkurthy, Steiner, Fang, Bai, and Chintala]{NEURIPS2019_9015}
Adam Paszke, Sam Gross, Francisco Massa, Adam Lerer, James Bradbury, Gregory
  Chanan, Trevor Killeen, Zeming Lin, Natalia Gimelshein, Luca Antiga, Alban
  Desmaison, Andreas Kopf, Edward Yang, Zachary DeVito, Martin Raison, Alykhan
  Tejani, Sasank Chilamkurthy, Benoit Steiner, Lu~Fang, Junjie Bai, and Soumith
  Chintala.
\newblock Pytorch: An imperative style, high-performance deep learning library.
\newblock In H.~Wallach, H.~Larochelle, A.~Beygelzimer, F.~d\textquotesingle
  Alch\'{e}-Buc, E.~Fox, and R.~Garnett, editors, \emph{Advances in Neural
  Information Processing Systems 32}, pages 8024--8035. Curran Associates,
  Inc., 2019.
\newblock URL
  \url{http://papers.neurips.cc/paper/9015-pytorch-an-imperative-style-high-performance-deep-learning-library.pdf}.

\bibitem[Redmon et~al.(2016)Redmon, Divvala, Girshick, and
  Farhadi]{redmon2016look}
Joseph Redmon, Santosh Divvala, Ross Girshick, and Ali Farhadi.
\newblock You only look once: Unified, real-time object detection, 2016.

\bibitem[Ren et~al.(2015)Ren, He, Girshick, and Sun]{NIPS2015_5638}
Shaoqing Ren, Kaiming He, Ross Girshick, and Jian Sun.
\newblock Faster r-cnn: Towards real-time object detection with region proposal
  networks.
\newblock In C.~Cortes, N.~D. Lawrence, D.~D. Lee, M.~Sugiyama, and R.~Garnett,
  editors, \emph{Advances in Neural Information Processing Systems 28}, pages
  91--99. Curran Associates, Inc., 2015.
\newblock URL
  \url{http://papers.nips.cc/paper/5638-faster-r-cnn-towards-real-time-object-detection-with-region-proposal-networks.pdf}.

\bibitem[Russakovsky et~al.(2014)Russakovsky, Deng, Su, Krause, Satheesh, Ma,
  Huang, Karpathy, Khosla, Bernstein, Berg, and
  Fei-Fei]{russakovsky2014imagenet}
Olga Russakovsky, Jia Deng, Hao Su, Jonathan Krause, Sanjeev Satheesh, Sean Ma,
  Zhiheng Huang, Andrej Karpathy, Aditya Khosla, Michael Bernstein,
  Alexander~C. Berg, and Li~Fei-Fei.
\newblock Imagenet large scale visual recognition challenge, 2014.

\bibitem[Schawinski et~al.(2017)Schawinski, Zhang, Zhang, Fowler, and
  Santhanam]{Schawinski_2017}
Kevin Schawinski, Ce~Zhang, Hantian Zhang, Lucas Fowler, and Gokula~Krishnan
  Santhanam.
\newblock Generative adversarial networks recover features in astrophysical
  images of galaxies beyond the deconvolution limit.
\newblock \emph{Monthly Notices of the Royal Astronomical Society: Letters},
  page slx008, Jan 2017.
\newblock ISSN 1745-3933.
\newblock \doi{10.1093/mnrasl/slx008}.
\newblock URL \url{http://dx.doi.org/10.1093/mnrasl/slx008}.

\bibitem[Simonyan and Zisserman(2014)]{simonyan2014deep}
Karen Simonyan and Andrew Zisserman.
\newblock Very deep convolutional networks for large-scale image recognition,
  2014.

\bibitem[Tonry et~al.(2018)Tonry, Denneau, Heinze, Stalder, Smith, Smartt,
  Stubbs, Weiland, and Rest]{Tonry_2018_2}
J.~L. Tonry, L.~Denneau, A.~N. Heinze, B.~Stalder, K.~W. Smith, S.~J. Smartt,
  C.~W. Stubbs, H.~J. Weiland, and A.~Rest.
\newblock Atlas: A high-cadence all-sky survey system.
\newblock \emph{Publications of the Astronomical Society of the Pacific},
  130\penalty0 (988):\penalty0 064505, May 2018.
\newblock ISSN 1538-3873.
\newblock \doi{10.1088/1538-3873/aabadf}.
\newblock URL \url{http://dx.doi.org/10.1088/1538-3873/aabadf}.

\end{thebibliography}

\end{document}